\documentclass[conference,letterpaper]{IEEEtran}
\IEEEoverridecommandlockouts

\usepackage[utf8]{inputenc}
\usepackage{amsmath,amssymb}
\usepackage{hyperref}
\usepackage{amsmath,amssymb,amsthm}
\usepackage{enumerate}
\usepackage{stfloats}
\usepackage{comment}
\usepackage{bbm}
\usepackage{subcaption}
\usepackage{siunitx}
\usepackage{mathtools}
\usepackage{accents,latexsym,cancel}
\usepackage{cite}

\usepackage[dvipsnames]{xcolor}
\definecolor{myPink}{RGB}{255,105,183}

\usepackage[T1]{fontenc}
\usepackage{graphics} 
\usepackage{epsfig} 
\usepackage[mathscr]{euscript}
\usepackage{algorithm}
\usepackage[noend]{algpseudocode}
\usepackage{bbm}
\makeatletter
\def\BState{\State\hskip-\ALG@thistlm}
\makeatother

\usepackage{tikz}
\usetikzlibrary{arrows,shapes,chains,matrix,positioning,scopes,patterns,calc,
decorations.markings,
decorations.pathmorphing,
}

\usepackage{pgfplots}
\pgfplotsset{compat=1.3}
\usepgflibrary{shapes}

\newtheorem{theorem}{Theorem}

\newtheorem{definition}[theorem]{Definition}

\newtheorem{remark}[theorem]{Remark}

\renewcommand{\epsilon}{\varepsilon}

\newcommand{\RNum}[1]{\uppercase\expandafter{\romannumeral #1\relax}}

\newcommand{\av}{\ensuremath{\mathbf{a}}}

\newcommand{\cv}{\ensuremath{\mathbf{c}}}

\newcommand{\hv}{\ensuremath{\mathbf{h}}}

\newcommand{\mv}{\ensuremath{\mathbf{m}}}
\newcommand{\nv}{\ensuremath{\mathbf{n}}}
\newcommand{\pv}{\ensuremath{\mathbf{p}}}
\newcommand{\qv}{\ensuremath{\mathbf{q}}}
\newcommand{\rv}{\ensuremath{\mathbf{r}}}
\newcommand{\sv}{\ensuremath{\mathbf{s}}}

\newcommand{\xv}{\ensuremath{\mathbf{x}}}

\newcommand{\Am}{\ensuremath{\mathbf{A}}}

\newcommand{\Ktot}{\ensuremath{K_{\mathrm{tot}}}}

\DeclareMathAlphabet{\mcl}{OMS}{cmsy}{m}{n}

\newlength\tikzwidth
\newlength\tikzheight

\definecolor{mycolor1}{rgb}{0.63529,0.07843,0.18431}%
\definecolor{mycolor2}{rgb}{0.00000,0.44706,0.74118}%
\definecolor{mycolor3}{rgb}{0.00000,0.49804,0.00000}%
\definecolor{mycolor4}{rgb}{0.87059,0.49020,0.00000}%
\definecolor{mycolor5}{rgb}{0.00000,0.44700,0.74100}%
\definecolor{mycolor6}{rgb}{0.74902,0.00000,0.74902}%

{\begin{list}{$\bullet$}
   {\setlength{\itemsep}{0in}
    \setlength{\labelsep}{0.05in}
    \setlength{\labelwidth}{0.2in}
    \setlength{\leftmargin}{0.25in}
    \setlength{\rightmargin}{0in}
    \setlength{\topsep}{0in}
   }}
{\end{list}}
\newcommand{\T}{^{\mbox{\tiny T}}}

\newcommand{\Pm}{\mathbf{P}}
\newcommand{\Wm}{\mathbf{W}}
\newcommand{\Qm}{\mathbf{Q}}
\newcommand{\Hm}{\mathbf{H}}
\newcommand{\Ym}{\mathbf{Y}}
\newcommand{\Xm}{\mathbf{X}}
\newcommand{\Zm}{\mathbf{Z}}

\newcommand{\Fm}{\mathbf{F}}
\newcommand{\pmd}{\text{P}_{\mathrm{md}}}
\newcommand{\pfa}{\text{P}_{\mathrm{fa}}}
\newcommand{\pe}{\ensuremath{\text{P}_{\mathrm{e}}}}
\title{FASURA: A Scheme for Quasi-Static Massive MIMO Unsourced Random Access Channels}
\author{Michail Gkagkos, Krishna R. Narayanan, Jean-Francois Chamberland, Costas N. Georghiades \\
Department of Electrical and Computer Engineering, Texas A\&M University
\thanks{
This material is based upon work supported, in part, by the National Science Foundation (NSF) under Grant CCF-2131106 and by Qualcomm Technologies, Inc., through their University Relations Program.}
}
\begin{document}

\maketitle

\begin{abstract}
This article considers the massive MIMO unsourced random access problem on a quasi-static Rayleigh fading channel.
Given a fixed message length and a prescribed number of channel uses, the objective is to construct a coding scheme that minimizes the energy-per-bit subject to a fixed probability of error.
The proposed scheme differs from other state-of-the-art schemes in that it blends activity detection, single-user coding, pilot-aided and temporary decisions-aided iterative channel estimation and decoding, minimum-mean squared error (MMSE) estimation, and successive interference cancellation (SIC).
We show that an appropriate combination of these ideas can substantially outperform state-of-the-art coding schemes when the number of active users is more than 100, making this the best performing scheme known for this regime.

\end{abstract}


\section{Introduction}


A new perspective on random access was proposed in \cite{yuryAWGN} to accommodate the type of traffic generated by unattended devices.
 Also, an extension of \cite{yuryAWGN}, when the number of users is unknown at the access point, is studied in \cite{GiuseppeUnknownK}.
The ensuing model, coined \emph{unsourced random access} (URA) in \cite{avinash2017}, has been widely adopted as a common task framework for emerging IoT wireless networks.
A motivation behind this model is that, as the prospective user population grows, the assignment of spectral resources based on queue lengths and channel conditions becomes impractical.
This is especially true when devices sporadically transmit short packets.
A pragmatic alternative is to have all active devices share a same codebook.
This way, the system can operate irrespective of the total user population, and target the number of active devices instead.
In such situations, the decoder aims to recover the set of messages regardless of user identities.
If a device wishes to reveal its identity, it can embed it in the payload of its own message.

Many conceptual approaches and candidate solutions tailored to URA have appeared in recent years, with a majority of them confined to receivers equipped with a single antenna.
These schemes are informed by the natural connection between URA and sparse support recovery in large dimensions.
Two noteworthy lines of work have emerged.
A first group of publications are inspired by compressed sensing (CS) solvers~\cite{VamsiCCSAWGN,SPARCAWGN,AlexeyRecoverableCodes}.
These schemes also leverage notions from forward error correction to enable the application of CS solvers to very large spaces.
A second set of results has been inspired by more traditional multiple access techniques, including notions from multi-user detection~\cite{VerduMUC} and random access.
Some of these scheme are currently the state-of-the-art for URA, depending on the parameters of operation~\cite{AsitPolar,AsitLDPC,IRSA,avinash2017}.

\subsection{Related Work} 

Many recent developments by the URA research community center on practical aspects of wireless communications, such as fading and MIMO models.
For instance, Andreev et al.\ in~\cite{AndreevPolarEM} examine a quasi-static fading URA channel, and they propose a scheme that blends polar codes and expectation-maximization (EM) clustering.
An early contribution to MIMO URA can be found in~\cite{VolodymyrBlessing}, where signals from different devices across slots are stiched together using channel properties rather than an outer code.
Fengler et al.\ in \cite{FenglerNonBay} combine a non-Bayesian sparse recovery algorithm with an outer code aimed at message disambiguation to facilitate MIMO URA; the performance of this algorithm can be improved using successive cancellation list decoding, as shown in \cite{amalladinne2021coded}.
Liu and Wang borrow ideas from slot-based transmissions, and they propose a receiver that merges simultaneous orthogonal matching pursuit (S-OMP) and bilnd channel estimation~\cite{LiuSparsity}. 
The same authors also propose an iterative receiver based on sparse Tanner graph; therein, each iteration can recover at most three codewords~\cite{LiuTanner}.
Sabulal and Bhashyam consider a special case of the MIMO URA problem, namely that of user activity detection, and propose a deep unfolding based algorithm in \cite{sabulal2020}.
 Cheng et al.\ solve the same problem using model-based algorithms in \cite{LiPingAMP_URA}.
In \cite{srivatsa2021IRSA}, Srivatsa and Murthy analyze the throughput of irregular repetition slotted aloha (IRSA), and  derive channel estimates for three schemes.
Decurninge et al.\ offer an efficient solution based on a tensor construction, with good performance in some regimes~\cite{MaximeTensor}.

Among published work, the scheme put forth by Fengler et al.\ offers the best performance~\cite{fengler2022pilotbased}.
This construction relies on preamble-selected pilot sequences to estimate channels, and message payloads are encoder using a polar code.
At the receiver, the algorithm exploit the fact that the recovery of pilot sequences can be modeled as a multiple measurement vector (MMV) problem, and they use approximate message passing (MMV-AMP) to recover the active set.
Then, channel coefficients are evaluated based on MMSE estimation and maximum ratio combining (MRC) is used to combine the signals from different antennas.
The outputs of the MRC serves as symbol estimates, and a polar decoder attempts to recover the most likely message.
Finally, a successive interference canceller subtracts the contribution of the decoded users, and the process starts anew with pilot recovery applied to the residual signal.

\subsection{Main Contributions}

Our proposed approach embraces lessons learned from previous schemes~\cite{avinash2017,AsitPolar}, 
and extend them to the MIMO URA setting.
Since our communication scheme incorporates random spreading, and it is designed for fading URA, we call it FASURA (fading spread unsourced random access). 
Like many previous contributions, it divides messages into two parts.
The preamble is dedicated to the selection of operational parameters, whereas the payload is encoded using a single-user code.
This mimics the architectures of \cite{IRSA,AsitPolar} for SISO systems, and \cite{fengler2022pilotbased} in the MIMO case.
Departures from \cite{fengler2022pilotbased} include the use of spreading sequences, the detection of active sequences, and distinct channel/symbol estimation techniques.
Specifically, instead of having devices send modulated polar codewords directly, 
FASURA spreads every coded symbol before transmission.
The idea is that forming a linear minimum-mean squared error (LMMSE) estimate of the coded symbols based on the active spreading sequences can mitigate interference during the decoding process, especially when the number of active devices is large.
In addition, we propose a new channel estimation step, which we call noisy pilot channel estimation (NOPICE).
Therein, the channel is estimated using both pilots and preliminary decisions about coded symbols.
Although, some preliminary symbol decisions are erroneous, the overall impact of this approach is better performance.
This estimation technique is embedded in an iterative loop and, when parameters are picked judiciously, progressively leads to better channel estimates.
With these two innovations, the proposed architecture outperforms the scheme put forth by Fengler et al.~\cite{fengler2022pilotbased} over a range of parameters.
For example, for $500$ active devices, the difference in ${E_{\mathrm{b}}}/{N_0}$ is more than $2.6$~dB.
This makes FASURA the state-of-the-art for MIMO URA in certain practical regimes.

\subsection{Notation}

Throughout, $\mathbb{Z}_+$ and $\mathbb{C}$ refer to the non-negative integers and complex numbers, respectively.
We use $[n]$ to denote $\{1,2,\dots,n\}$.
We employ boldface lowercase $\av$ and boldface uppercase letters $\Am$ to indicate vectors and matrices.
Sets are labeled with calligraphic letters, e.g., $\mathcal{A}$.
We also adopt programming style notation with $\Am[:,t]$ and $\Am[k,:]$ representing the $t$th column and $k$th row of $\Am$.

\section{System Model}
\label{st:system}
Consider the uplink of a wireless network with a total of $\Ktot$ devices out of which $K$ devices are active $(K \ll \Ktot)$.
For convenience, we label active users using integers from the set ${\cal K}=[K]$.
All the active users share the same $n$ complex channel uses, and each of them wishes to transmit a $B$-bit message to a common destination.
Every user is equipped with a single antenna, whereas the access point features $M$ antennas.
We consider a quasi-static Rayleigh fading model whereby channel coefficients remain fixed during the transmission of an entire codeword.
Furthermore, the antennas at the access point are located far enough from one another as to create independent channel realizations.

Let $\mv_k$ be $B$-bit message of user~$k$, and $\xv_k = \mathcal{E}(\mv_k) \in \mathbb{C}^n$ be the encoded and modulated signal (input to the channel) corresponds to the message  $\mv_k$.
Then, the received signal takes the form
\begin{equation} \label{eq:systemSum}
\begin{split}
    \Ym &= \sum_{k=1}^{\Ktot} \delta_k \xv(\mv_k) \hv_k\T + \Zm
    = \sum_{k \in {\cal K}} \xv(\mv_k) \hv_k\T + \Zm ,
\end{split}
\end{equation}
where $\delta_k$ is an indicator function that takes value one when user~$k$ is active, and zero otherwise.
Vector $\hv_k \in \mathbb{C}^M$ captures the channel coefficients between user~$k$ and the $M$ receive antennas.
The elements of $\hv_k$ are independent complex Gaussian random variables with mean zero and unit variance, as prescribed by Rayleigh fading.
Additive noise component $\Zm \in \mathbb{C}^{n \times M}$ is a vector with i.i.d.\ entries, each drawn from a circularly symmetric complex Gaussian distribution ${\cal CN}(0,\sigma_z^2)$.
Furthermore, every transmit signal must satisfy power constraint $\| \xv(\mv_k) \|^2 \leq P$; for simplicity, we assume that $P = 1$.
As a result, we can define the energy per bit to noise power spectral density ratio of the system by
\begin{equation*}
\frac{E_{\mathbf{b}}}{N_0} = \frac{\Vert \xv(\mv_k)\Vert^2}{B\sigma_z^2}. 
\end{equation*}
At the access point, the decoder aims to produce a set $\hat{{\cal K}}$ of candidate messages with cardinality at most $K$.
The system performance is evaluated in terms of the probability of missed detection $\pmd$ and probability of false alarm  $\pfa$,
\begin{xalignat*}{2}
\pmd &= \frac{\mathbb{E}[n_{\mathrm{ms}}]}{K} & \pfa &= \mathbb{E}\left[ \frac{n_{\mathrm{fa}}}{\hat{K}} \right]
\end{xalignat*}
where $n_{\mathrm{ms}}$ and $n_{\mathrm{fa}}$ denote the number of misses and false alarms, respectively.
Variable $\hat{K}$ represents the number of messages declared by the decoder.
We note that the expectation is taken over the randomness of the fading process, the noise process, and the relevant algorithmic components.
We define the probability of error $\pe$ to be the sum of the two types of probability introduced above,
$\pe = \pmd + \pfa$.
For fixed parameters $B$, $n$, $K$, $M$, and target error probability $\epsilon$, our objective is to construct a communication scheme that minimizes ${E_{\mathrm{b}}}/{N_0}$ while also satisfying the constraint  $\pe \leq \epsilon$.

\section{FASURA}
\label{sc:fasura}

We proceed with the description of our proposed scheme, FASURA.
We begin with an overview of the encoding process, and then we discuss our decoding strategy.

\subsection{Encoder}

Active user~$k$ aims to transmit message $\mv_k$ to the destination.
Following the URA philosophy, encoding is dissociated from the identity of the user, and the sent signal only depends on the content of $\mv_k$.
As such, we describe the encoding process for a generic message $\mv$, with the understanding that the steps are repeated by all the active devices.
The encoder first splits the message into two parts $\mv_f$ and $\mv_s$ of lengths $B_f$~bits and $B_s$~bits, respectively.
The two parts $\mv_f$ and $\mv_s$ are encoded as follows.

\subsubsection{Encoding $\mv_f$}
Let ${\cal A} = \{ \Am_t \}_{t=1}^T$ denote a collection of $T=n_c/2$ spreading matrices indexed by time $t$.
{The coefficients of matrix $\Am_t$, are distributed as complex Gaussian random variables, i.e. $a_{t,j} \sim {\cal CN}(0,1)$.
The columns of matrix $\Am_t$ can be viewed as spreading sequences of length $L$, which can be utilized at time~$t$.}
The entries of $\Am_t$ are normalized to have energy of  $1/2n$ since they will be used in conjunction with Quadrature Phase Shift Keying (QPSK) symbols chosen from $\{\pm 1 \pm j\}$, which are normalized to have an energy of 2. The product is then normalized to have unit energy.
There are $J = 2^{B_f}$ possible spreading sequences attached to every time instant.
Likewise, let ${\bf P} \in \{ \pm \frac{1}{\sqrt{2n}} \pm \frac{j}{\sqrt{2n}} \}^{n_p \times J}$ be a matrix whose columns are possible pilot sequences.
The selection of spreading sequences and pilots is performed using $\phi : \{0,1\}^{B_f} \rightarrow [J]$, a bijection that maps $\mv_f$ to an index in $[J]$.
Thus, when the first part of the message to be sent is $\mv_f$, then the corresponding user employs the series of spreading sequences $\{ \Am_t[:,\phi(\mv_f)]\}_{t=1}^T$, along with pilot sequence $\Pm[:,\phi(\mv_f)]$.
Thus, the overall encoding function for $\mv_f$ can be summarized as follows,
\begin{equation*}
    g(\mv_f) \rightarrow \{ \Am_t[:,\phi(\mv_f)]\}_{t=1}^T \cup \{ \Pm[:,\phi(\mv_f)] \}.
\end{equation*}


An important aspect of the proposed scheme is that $J \ll \Ktot$, which is critical in limiting the complexity at the decoder.
We emphasize that there are no guarantees that every active user will select a distinct sequence from an orthogonal set.
Rather, active users pick sequences at random from a non-orthogonal set.
In fact, the URA framework makes it impossible to hand pick sequence since two devices with a common message will ultimately transmit the same signal.

\subsubsection{Encoding $\mv_s$}

The second part of the message, namely $\mv_s$, is first encoded using a cyclic redundancy check (CRC) code.
The resulting codeword of length $B_{c} = B_s + B_{crc}$ then acts as input to an encoder for a $(n_c,B_c)$ polar code with $n_c-B_c$ frozen bit positions.
We construct at random a matrix $\Fm \in \{0,1\}^{n_c - B_{c} \times J}$, and the frozen bits for an active user are chosen to be the $\phi(\mv_f)$th column of $\Fm$. 
In effect, this corresponds to each user choosing a coset of a polar code depending on $\phi(\mv_f)$.
This step provides an additional level of error detection to identify false alarms when estimating the active spreading sequences at the receiver.
Suppose $\cv \in \{0,1\}^{n_c}$ is the output of the polar encoder, and let ${\cal{I}} = \{\pi_1(\cdot), \ldots, \pi_J(\cdot)\}$ be a set of $J$ interleavers.
The polar codeoword $\cv$ is permuted using $\pi_{\phi(\mv_f)}(\cdot)$, and the ensuing vector $\tilde{\cv} = \pi_{\phi(\mv_f)}(\cv)$ is modulated using QPSK to obtain vector $\sv$ of length $n_c/2$.
Finally, the $t$th symbol of $\sv$ is spread using the $\phi(\mv_f)$th column of $\Am_t$.
The resulting signal $\qv$ can be expressed as 
\begin{equation} \label{eq:secondPart}
    \qv(\mv_f,\mv_c) = \begin{bmatrix} s_1\av_1\T  & s_2\av_2\T  & \cdots & s_T\av_T\T \ \end{bmatrix}\T
\end{equation}
where $s_t$ is a QPSK symbol and $\av_t = \Am_t[:,\phi(\mv_f)]$ is the $\phi(\mv_f)$th column of $\Am_t$.
The input signal to the channel is the concatenation of the pilot sequence $\pv(\mv_{k,f})$ and spread codeword $\qv(\mv_f,\mv_c)$. 
Altogether, when the message of user~$k$ is $\mv_k = \left( \mv_{k,f}, \mv_{k,c} \right)$, 
the signal sent by this user is equal to
\begin{equation*}
\xv_k = \begin{bmatrix} 
\pv\T(\mv_{k,f}) &
\qv\T(\mv_{k,f},\mv_{k,c}) \end{bmatrix}\T
\end{equation*}
where $\pv(\mv_{k,f}) = \Pm[:, \phi(\mv_{k,f})]$ and note  that $\Vert \xv \Vert^2 = 1$.
With this procedure, the system model of \eqref{eq:systemSum} can be written as the concatenation of
\begin{xalignat}{2} \label{eq:Yp}
\Ym_p &=  \Pm_a \Hm_a + \Zm_p\quad\quad  \text{and}&
\Ym_q &= \Qm_a\Hm_a + \Zm_q
\end{xalignat}
or, in vector form,
\begin{equation*}
\begin{bmatrix} \Ym_p \\ \Ym_q \end{bmatrix}
= \begin{bmatrix}  \Pm_a \\
\Qm_a \end{bmatrix} \Hm_a
+ \begin{bmatrix} \Zm_p \\ \Zm_q \end{bmatrix}
\end{equation*}
where subscript $a$ indicates sub-matrices with active columns only.
That is, the $k$th column of $\Pm_a$ is $\Pm[:,\phi(\mv_{k,f})]$ and the $k$th column of $\Qm_a$ is $\qv_k$.
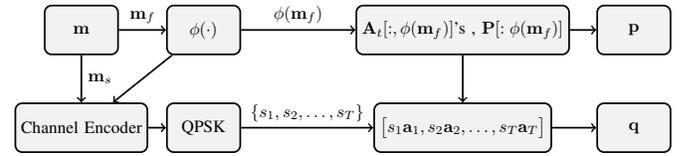
\begin{figure}[h]
    \centering
    \scalebox{0.65}{\begin{tikzpicture}
  [
  font=\normalsize, draw=black, line width=1pt,
  block/.style={rectangle, minimum height=10mm, minimum width=15mm,
  draw=black, fill=gray!10, rounded corners},
  ED/.style={rectangle, minimum height=10mm, minimum width=15mm, draw=black, rounded corners},
  message/.style={rectangle, minimum height=5.5mm, minimum width=15mm, draw=black, rounded corners}
  ]

\node[block,align=center] (message)  {$\mv$};

\node[block,right of= message, node distance = 2.5cm] (hash) {$\phi(\cdot)$};
\draw[->] (message) -- (hash) node[midway,sloped,above] {$\mv_f$};

\node[block,right of= hash, node distance = 5.3cm] (matrices)  {$\Am_t[:,\phi(\mv_f)]$'s , $\Pm[:\phi(\mv_f)]$};
\draw[->] (hash) -- (matrices)node[midway,sloped,above] {$\phi(\mv_f)$};;

\node[block,below of= message, node distance = 2cm] (encoder){Channel Encoder};
\draw[->] (message) -- (encoder) 
node[midway,sloped,right,rotate=90]{$\mv_s$};
\draw[->] (hash) -- (encoder);

\node[block,below of= hash, node distance = 2cm] (qpsk){QPSK};
\draw[->] (encoder) -- (qpsk);

\node[block,below of= matrices, node distance = 2cm] (spread){\big[$s_{1}\av_1,s_{2}\av_2,\dots,s_{T}\av_T\big]$};
\draw[->] (qpsk) -- (spread)node[midway,sloped,above] {$\{s_1,s_2,\dots,s_T\}$};
\draw[->] (matrices) -- (spread);

\node[block,right of= matrices, node distance = 3.5cm] (p){$\pv$};

\node[block,below of= p, node distance = 2cm] (q){$\qv$};

\draw[->] (matrices) -- (p);
\draw[->] (spread) -- (q);

\end{tikzpicture}}
    \caption{This block diagram offers a synopsis of the encoding process.}
    \label{fig:encoder}
\end{figure}
\subsection{Decoder}
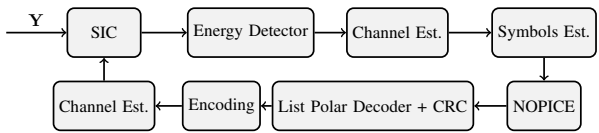
\begin{figure}[h]
    \centering
    \scalebox{0.65}{\begin{tikzpicture}
  [
  font=\normalsize, draw=black, line width=1pt,
  block/.style={rectangle, minimum height=10mm, minimum width=15mm,
  draw=black, fill=gray!10, rounded corners},
  sum/.style={rectangle, minimum height=15mm, minimum width=15mm, ddraw=black, rounded corners, fill=gray!10},
  message/.style={rectangle, minimum height=5.5mm, minimum width=15mm, draw=black, rounded corners}
  ]

\node[block,align=center] (sic)  {SIC};
\node[block,right of= sic, node distance = 3cm] (ed) {Energy Detector};
\draw[->] (sic) -- (ed);

\node[block,right of= ed, node distance = 3cm] (chEst) {Channel Est.};
\draw[->] (ed) -- (chEst);

\node[block,right of= chEst, node distance = 3cm] (syEst) {Symbols Est.};
\draw[->] (chEst) -- (syEst);

\node[block,below of= syEst, node distance = 1.5cm] (nopice) {NOPICE};
\draw[->] (syEst) -- (nopice);

\node[block,left of = nopice, node distance = 3.5cm] (dec) {List Polar Decoder + CRC};
\draw[->] (nopice) -- (dec);

\node[block,left of= dec, node distance = 3.1cm] (enc) {Encoding};
\draw[->] (dec) -- (enc);

\node[block,below of= sic, node distance = 1.5cm] (chEst2) {Channel Est.};
\draw[->] (enc) -- (chEst2);

\draw[->] (chEst2) -- (sic);

\draw[->] (-2,0) --  (sic.west)   node[midway,above]{${\bf Y}$};

\end{tikzpicture}}
    \caption{This notional diagram outlines the message recovery process at the receiver.
    This iterative scheme includes the identification of selected spreading sequence, channel/symbol estimation, and polar decoding.}
    \label{fig:decoder}
\end{figure}

We construct an iterative receiver that aims to recover the messages transmitted by the active users.
Its main components are an energy detector, a series of MMSE estimators, and a list-polar decoder.
Because of the nature of the URA, the pilot sequences cannot be assigned to the users beforehand.
As a result, the first step in message recovery is to identify the active columns of $\Pm$ using an energy detector.
Then MMSE estimators are utilized to estimate the channel coefficients and the QPSK symbols transmitted by the active users.
Afterward, a polar list-decoder outputs the most likely messages.
Finally, a successive interference canceller subtracts the estimated contribution of the decoded messages, and the  procedure starts anew on the residual signal, iterating until the decoder cannot output a new message.
The main building blocks of the algorithm are highlighted in Fig.~\ref{fig:decoder}.
We elaborate on individual components below.

\subsubsection{Energy Detector}
An energy detector is used to determine active spreading sequences.
Mathematically, this is accomplished by correlating the received signal $\Ym$ with the corresponding columns in $\Pm$ and $\{ \Am_t \}_{t=1}^T$ and computing the statistic $\lambda_j$ given by
\begin{equation} \label{equation:EnergyDetector}
    \lambda_j = \Vert\Pm^*[:,j]\Ym_p\Vert^2 + \sum_{t=1}^T\Vert\Am_t^*[:,j]\Ym_q[\nv_t,:]\Vert^2, \forall j \in [J]
\end{equation}
where $\nv_t = [(t-1)L + 1 \mathpunct{:} tL]$.
The energy detector computes $\lambda_j$ for all $j \in  [J]$ and outputs the indices corresponding to the largest $K$ values. 
Let $\hat{\mathcal{M}}_f$ denote the set of $\mv_f$'s that correspond to the $K$ largest values of $\lambda_j$'s.

After this step, the active columns of $\Pm$ and $\{ \Am_t \}_{t=1}^T$, the values of the frozen positions, and the interleavers are considered known for the purpose of the algorithmic progression.
Thus, the decoder can move on to the recovery of the second part of every message.

\subsubsection{Channel Estimation}
Channel coefficients have to be estimated before proceeding with the symbol estimation step.
An MMSE filter can be derived to estimate the SIMO channels of the candidate users included in $\hat{{\cal K}}$.
Using active pilots from the first part of the received signal $\Ym_p$, as in \eqref{eq:Yp}, we obtain
\begin{align*}
    \Wm_1 = \left( {\bf I}_{\hat{K}} + \frac{\hat{\Pm}^*\hat{\Pm}}{\sigma_z^2} \right)^{-1}\frac{\hat{\Pm}^*}{\sigma_z^2}
\end{align*}
where $\hat{\Pm} = \Pm[:,\phi(\hat{\mathcal{M}}_f)]$.
Since the filter is independent of the antenna index, the estimated channel coefficients between the $\hat{K} = |\hat{{\cal K}}|$ users and the $M$ antennas are taken to be
\begin{align}
    \hat{\Hm} = \Wm_1 \Ym_p
\end{align}
where $\hat{\Hm}$ is a $\hat{K} \times M$ matrix containing all the estimated coefficients.

\subsubsection{Symbol Estimation}
Since the active columns of $\Am_t$'s and the channel coefficients have been estimated, the next step is to recover the QPSK symbols.
We can write the received signal (of length $L$) for each symbol $t$ as follows
\begin{equation*}
\begin{split}
\mathbf{Y}_q[\nv_t,m] &= \Am_t \operatorname{diag}(\rv_t) \Hm[:,m] + \mathbf{Z}[\nv_t,m] \\
&= \Am_t \operatorname{diag} \left( \mathbf{H}[:,m] \right) \rv_t + \mathbf{Z}[\nv_t,m]
\end{split}
\end{equation*}
where $\rv_t = \left( s_{t,1},s_{t,2},\ldots,s_{t,K}\right)$ are the symbols of the users at time $t$.
By stacking the columns of $\Ym_q$, we can obtain the expression below
\begin{equation*}
\begin{split}
\underbrace{\begin{bmatrix} 
\Ym_q[\nv_t,1] \\
\vdots \\
\Ym_q[\nv_t,M]
\end{bmatrix}}_{\mathtt{y}_t \in \mathbb{C}^{L M \times 1}}
&= \underbrace{\begin{bmatrix}
\Am_t \operatorname{diag} \left( \mathbf{H}[:,1] \right) \\
\vdots \\
\Am_t \operatorname{diag} \left( \mathbf{H}[:,M] \right)
\end{bmatrix}}_{\mathtt{B}_t \in \mathbb{C}^{L M \times K}} \; \underbrace{\rv_t}_{K \times 1} \;
+ \; \underbrace{\begin{bmatrix} \mathbf{Z}[\nv_t,1] \\
\vdots \\ 
\mathbf{Z}[\nv_t,M]
\end{bmatrix}}_{\mathtt{z}_t \in \mathbb{C}^{L M \times 1}} .
\end{split}
\end{equation*}
As a consequence, we can apply an MMSE-like estimator to the vectorized received signal
\begin{equation} \label{eq:vectForm}
    \mathtt{y}_t = \mathtt{B}_t \rv_t + \mathtt{z}_t .
\end{equation}
Since active spreading sequences and channel coefficients are not known a priori, we modify \eqref{eq:vectForm} to account for the estimation inaccuracies,
\begin{equation} \label{eq:vectFormError}
    \mathtt{y}_t = \hat{\mathtt{B}}_t\rv_t + \big(\mathtt{B}_t - \hat{\mathtt{B}}_t\big)\rv_t + \mathtt{z}_t .
\end{equation}
In view of the last equation, one could take into consideration the interference term and increase the effective noise variance.
However, based on our simulations, the performance of our scheme essentially remains unaffected when the second term is neglected and, as such, we retain the simpler form.
Disregarding the interference term, the MMSE filter for \eqref{eq:vectFormError} takes the form
\begin{equation*}
\mathtt{W}_t
= \left( 2{\mathtt I}_{\hat{K}} + \frac{\hat{\mathtt{B}}^*_t\hat{\mathtt{B}}_t}{\sigma_z^2} \right)^{-1}
\frac{\hat{\mathtt{B}}^*_t}{\sigma_z^2} .
\end{equation*}
We stress that the set of the spreading sequence changes over time, therefore an MMSE filter has to be computed for every $t \in [T]$.
The estimated symbols of the active users at time~$t$ are given by
\begin{equation*}
\hat{\rv}_t = \mathtt{W}_t \mathtt{y}_t, \ \forall \quad t \in [T] .
\end{equation*}

\subsubsection{NOPICE}
One of the salient features of this scheme is the estimation of the channel using, not only the original pilots, but also the temporary coded decisions.
\begin{definition}(Temporary Coded Decisions) \label{df:tcd}
Once step~3 above is complete, the symbols across time corresponding to the same user are aggregated in the form of noisy codewords.
These signals are then passed to a polar list-decoder.
(The channel decoder block is explained in the next section.)
The output of this decoder block is a set of most likely messages.
These messages are subsequently re-encoded and modulated.
We call the outcome of this process \emph{temporary coded decisions}.
\end{definition}

\begin{figure}[h]
    \centering
    \scalebox{0.65}{\begin{tikzpicture}
  [
  font=\normalsize, draw=black, line width=1pt,
  block/.style={rectangle, minimum height=10mm, minimum width=15mm,
  draw=black, fill=gray!10, rounded corners},
  ED/.style={rectangle, minimum height=10mm, minimum width=15mm, draw=black, rounded corners},
  message/.style={rectangle, minimum height=5.5mm, minimum width=15mm, draw=black, rounded corners}
  ]







\node[block,align=center] (dec)  {List Polar Decoder};
\draw[->] (-3.5,0) --  (dec.west)   node[midway,above]{Symbols};
\node[block,right of= dec, node distance = 3cm] (crc) {CRC};
\draw[->] (dec) -- (crc);

\node[block,right of= crc, node distance = 2.5cm] (encoder) {Encode $\hat{\mv}_s$};
\draw[->] (crc) -- (encoder);

\node[block,below of= encoder, node distance = 2.2cm] (P)
{$\Pm$};

\node[block,left of= P, node distance = 2.5cm] (chEst) {Channel Est.};
\draw[->] (P) -- (chEst) node[midway,sloped,above] {$\hat{\Pm}$};

\draw[->] (encoder) -- (chEst)node[midway,sloped,above,rotate=330] {$\hat{\Qm}$};

\node[block,left of= chEst, node distance = 3.5cm] (syEst) {Symbols Est.};
\draw[->] (chEst) --(syEst) node[midway,sloped,above] {$\hat{\Hm}$};

\draw[->]  (syEst.west) -- (-3.5,-2.2)   node[midway,above]{Symbols};

\end{tikzpicture}}
    \caption{This block diagram highlights the main functionalities of NOPICE.
    The channel is estimated using pilots and estimated symbols, although the latter may include errors.}
    \label{fig:nopice}
\end{figure}
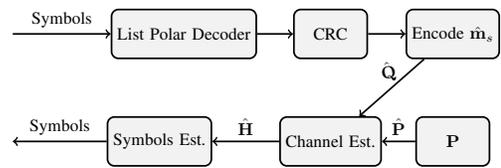

To produce such temporary coded decisions for all users, the first three blocks in NOPICE have to be a polar decoder, an encoder, and a modulator.
Let $\hat{\sv}_k = [\hat{s}_{1,k} \ \hat{s}_{2,k} \ \dots \ \hat{s}_{T,k}]$ be the temporary coded decisions of user $k$,
where $\hat{s}_{t,k}$ is the $t$th symbol of the user~$k$.
Since we know the pilots, the spreading sequences, and the symbols candidates for this user, we can construct its channel input,
\begin{equation*}
    \hat{\xv}_k =
    \begin{bmatrix}
    \pv\T (\hat{\mv}_{k,f})  & \qv\T(\hat{\mv}_{k,f},\hat{\mv}_{k,s})
    \end{bmatrix}\T
\end{equation*}
where $\hat{\qv}_k$ can be constructed as \eqref{eq:secondPart}, by replacing the true values with the candidate symbols.
Our strategy is to have $\hat{\xv}_k$'s act as a known signal while re-estimate the channel coefficients.
We stress that some of the symbols in $\hat{\sv}_k$ may be erroneous.
Nevertheless, they are used to re-estimate the channel in our algorithm, with the hope that most candidate symbols are correct.
Consider the received signal at antenna~$m$.
\begin{equation*}
\Ym[:,m] = \hat{\Xm}\Hm[:,m] +  \left(\Xm -\hat{\Xm} \right) \Hm[:,m] + \Zm[:,m]
\end{equation*}
By ignoring the interference term, we can apply the following MMSE-like filter to estimate the channel coefficients,
\begin{equation}
    \Wm_2 = \hat{\Xm}^* \left(\hat{\Xm}^*\hat{\Xm} + \mathbf{I}_{\hat{K}} \right)^{-1} \label{eq:mmse2}
\end{equation}
Once more, the MMSE filter is independent of the antenna index and, hence, $\Wm_2$ can be applied directly to observation $\Ym$.
At this stage, the updated channel estimates are used to re-estimate the QPSK symbols by the procedure described before.
Figure~\ref{fig:nopice} shows a block diagram of this process.

\subsubsection{Channel Decoder}
This block consists of a single-user list polar decoder, and a CRC validation step.
We describe the process for user~$k$, with the understanding that these step are reproduced for all the candidate users in $\hat{{\cal K}}$.
We note that, as a consequence, part of this procedure can be parallelized in a straightforward manner.
As mentioned before, the symbol estimates and the frozen values $\Fm[:,\phi(\hat{\mv}_f)]$ are passed to the decoder.
After completion, the polar decoder provides a list of  $n_L$ likely messages.
CRC validation is applied to elements of this list, and consistent messages that meet the CRC structure are retained.
The most likely message within the pruned list is returned by this block.

\begin{remark}(List-Decoder and NOPICE)
Under normal operation, the polar/CRC decoder returns the most likely consistent message.
Yet, it is possible that the list contains no codewords that fulfill their CRC constraints.
In this case, the most likely and, necessarily, inconsistent message is returned.
The hope is that, although the message is guaranteed to be wrong, a portion of the encoded symbols can still be correct.
\end{remark}

\subsubsection{SIC}
The final step in this composite iterative algorithm seeks to mitigate interference through successive interference cancellation (SIC).
This is accomplished by calculating the estimated channel input $\hat{\xv}$ for all the users whose most likely messages met the CRC validation process.
Then a channel estimation similarly to \eqref{eq:mmse2} is done.
Finally, we subtract their contribution from the received signal and pass the residual to the energy detector.
\begin{remark}(Residual)
When the first iteration has passed, all the building blocks of the receiver use the residual.
Nevertheless, SIC uses the received signal and produces a new residual for the next iteration.
\end{remark}

\section{Simulation Results}
\label{sc:sim}
We compare FASURA\footnote{The source code for the FASURA communication scheme is available at \url{https://github.com/EngProjects/mMTC}.} with the state-of-the-art scheme proposed by Fengler et al.~\cite{fengler2022pilotbased} to assess its performance.
To ensure a fair comparison between the two URA communication schemes, we pick parameters for our system that match their reported implementation.
Specifically, we choose $B=100$ message bits, $n=3200$ complex channel uses, and a target probability of error $\pe \leq 0.05$.
Beyond these constraints, the other parameters for FASURA are $n_p = 896$, $L = 9$, $n_c = 512$, $n_L = 64$ and $J = 2^{16}$.
We randomly generate $T = \frac{n_c}{2} = 256$ spreading sequence matrices and one pilot matrix.
{ Elements of the spreading sequence matrices are drawn independently from a complex Gaussian distribution with zero mean and variance 1.
Then the entries are normalized to have energy $1/2n$.}
Similarly, elements of the pilot matrix are generated independently and with equal probability from
$\left\{\pm \frac{1}{\sqrt{2n}} \pm \frac{j}{\sqrt{2n}} \right\}$.
For {$K = 100$, we use 12 CRC bits; whereas when $K > 100$, we employ 16 CRC bits}.
The number of antennas at the base station is set to $M=50$.

\begin{figure}
\centering
  \begin{tikzpicture}
\definecolor{mycolor1}{rgb}{0.63529,0.07843,0.18431}%
\definecolor{mycolor2}{rgb}{0.00000,0.44706,0.74118}%
\definecolor{mycolor3}{rgb}{0.00000,0.49804,0.00000}%
\definecolor{mycolor4}{rgb}{0.87059,0.49020,0.00000}%
\definecolor{mycolor5}{rgb}{0.00000,0.44700,0.74100}%
\definecolor{mycolor6}{rgb}{0.74902,0.00000,0.74902}%
\definecolor{mycolor7}{rgb}{0.502,0.2000,0.5902}

\begin{axis}[
font=\footnotesize,
width=7cm,
height=6cm,
scale only axis,
xmin=100,
xmax=800,
xtick = {100,200,...,800},
xlabel={\small Number of active users $K$},
xmajorgrids,
ymin=-12.35,
ymax=-5.75,
ytick = {-6,-7,...,-12},
ylabel={\small Required $E_{\mathrm{b}}/N_0$ (dB)},
ylabel near ticks,
ymajorgrids,
legend style={font=\footnotesize, at={(1, 1.3)},anchor=north east, draw=black,fill=white,legend cell align=left}
]

\addplot [color=mycolor2,solid,mark=diamond,line width=2.0pt]
  table[row sep=crcr]{%
10	-6.5\\
25	-6.5\\
50	-6.5\\
100	-7\\
200	-7\\
300	-6.5\\
500	-6\\
650	-6\\
700	35\\
};
\addlegendentry{TBM (Decurninge et al.~\cite{MaximeTensor}), $\ensuremath{\text{P}_{\mathrm{e}}} = 0.1$};

\addplot [color=mycolor1,solid,line width=2.0pt,mark size=1.4pt,mark=o,mark options={solid}]
  table[row sep=crcr]{
100	-11.97265625\\
200	-11.1474609375\\
300	-10.2587890625\\
400	-9.1162109375\\
500	-7.8466796875\\
600	-6.0693359375\\
700	-3.2763671875\\
800	1.2939453125\\
};
\addlegendentry{Pilot-Based (Fengler et al.~\cite{fengler2022pilotbased}), $\ensuremath{\text{P}_{\mathrm{e}}} = 0.05$};

\addplot [color=mycolor3,solid,line width=1.5pt,mark size=1.0pt,mark=square,mark options={solid}]
  table[row sep=crcr]{
100	 -12\\
200	-11.65\\
300	-11.25\\
400 -10.80	\\
500	-10.35\\
600 -9.65\\
700 -8.85\\
800 -7.7\\
};
\addlegendentry{FASURA without NOPICE, $\ensuremath{\text{P}_{\mathrm{e}}} = 0.05$};

\addplot [color=mycolor6,solid,mark=square,line width=1.5pt]
  table[row sep=crcr]{
100	-12.3\\
200	-12\\
300	-11.65\\
400	-11.30\\
500	-10.90\\
600 -10.20\\
700 -9.5\\
800 -8.5\\
};
\addlegendentry{FASURA, $\ensuremath{\text{P}_{\mathrm{e}}} = 0.05$};

\end{axis}

\end{tikzpicture}%
  \caption{This plot compares the performance of the proposed scheme to the performance of previously published schemes.
  The number of antennas at the base station is $M = 50$, users each wish to transmit $B=100$ bits of information, and the total number of channel uses is $n=3200$.
  The target probability of error is set to $\pe \leq 0.05$.
  The proposed scheme outperforms the state-of-the-art.}
  \label{fig:results}
\end{figure}
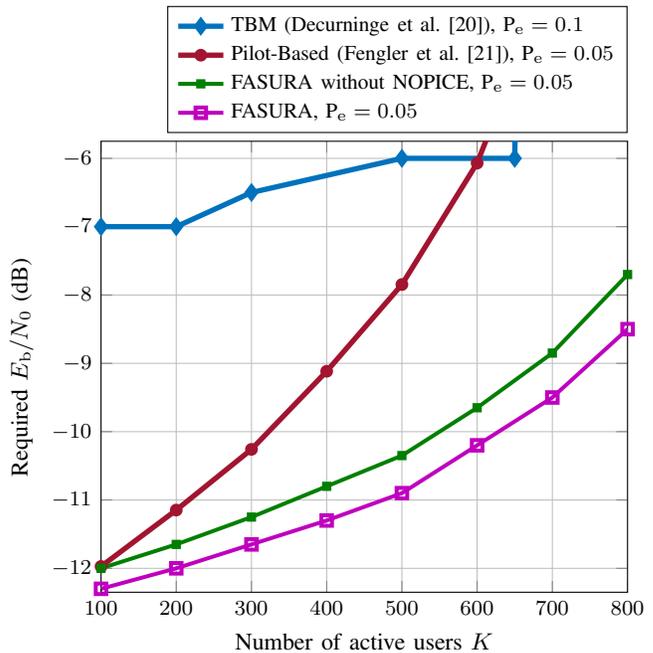

Figure~\ref{fig:results} plots the performance of FASURA, along with that of the communication scheme by Fengler et al.\ found in \cite{fengler2022pilotbased}.
Also, to motivate the use of the NOPICE block, we report the performance of FASURA with and without the NOPICE channel estimation technique.
For the operational parameters studied and a user population exceeding 100 active devices, FASURA outperforms the scheme proposed in~\cite{fengler2022pilotbased}.
Interestingly, as the number of active users grows, the gap between the two schemes widens.
{For example, when the number of users goes from $100$ to $800$, the gap increases from $0.3$~dB to $9$~dB.}
Furthermore, the presence of the NOPICE block seems to uniformly improve the performance of FASURA.
Our proposed scheme also substantially outperforms the tensor based modulation scheme in \cite{MaximeTensor}.
We should mention, however, that these benefits in terms of $E_{\mathrm{b}}/N_0$ come at the expense of additional computations.



\section{Conclusion}
This article considers the massive MIMO unsourced random access problem on a quasi-static Rayleigh fading channel.
We propose a novel communication scheme called FASURA that outperforms existing schemes when $K \geq 100$.
FASURA splits the payload into two parts and encodes the first part to a set of randomly generated pilots and spreading sequences.
The remainder of the information bits are encoded using a polar code.
Then the modulated coded symbols are spread using the spreading sequences.
The receiver is equipped with multiple antennas, and the objective is to recover the transmitted messages.
The decoding process includes; pilot detection, channel and symbol estimation, polar list decoding, and successive interference cancellation.
We explore a different way to perform channel estimation, which we call NOPICE, whereby after temporary decoding decisions have been made, the channel is re-estimated assuming that the decoded messages are accurate.
This scheme is somewhat reminiscent of the \emph{certainty equivalent principle} in control theory.
With NOPICE, the proposed scheme outperforms the pilot-based scheme recently published in \cite{fengler2022pilotbased}, when the number of users is more than $100$.
Also, through numerical simulations, we conduct a comparative study of FASURA with and without NOPICE.
{ It seems that the approach adopted within the NOPICE block leads to uniformly better performance, with gains of $0.4$~dB on average.}

\section{Future Work}
While the results for FASURA are encouraging, the parameters of the scheme have not been optimized, and it may be interesting to see how much performance we can gain through fine turning.
The tradeoff between the length of the spreading sequence and code rate should be investigated.
An interesting question to raise is when the spreading sequence plays an important role to mitigate the interference during the single-user decoding when the number of antennas is fixed.
From a design point of view, a key question would be if there are spreading sequences with complex entries that have better correlation properties.
Furthermore, by introducing NOPICE,  a question arises on how many pilots should be allocated in order to start the decoding process.
Another future task is to investigate when the NOPICE block fails.
In other words, how many errors can the system handle when the channel is re-estimated using the pilots and the temporary decisions.


Very recently, a scheme based on using multiple stages of orthogonal pilots has been proposed in \cite{TolgaOrthogonal2022} and it has been shown to exhibit good performance. A comprehensive comparison between our proposed scheme and the scheme in \cite{TolgaOrthogonal2022} would illuminate the advantages and disadvantages of spreading versus sparsification in time.

\bibliographystyle{IEEEbib}
\bibliography{references}

\end{document}